\documentstyle[11pt]{article}

\evensidemargin=\oddsidemargin
\begin{document}
\begin{flushright}
{\sf  VPI-IHEP-94-04~,~~
      TUIMP-TH-94/60~,~~ MSUHEP-40909 }\\
{September, 1994} \\
\end{flushright}
\begin{center}
{\Large
 \bf
            EQUIVALENCE THEOREM AND PROBING THE \\
            ELECTROWEAK SYMMETRY BREAKING SECTOR   }\\[1.3cm]

{\bf Hong-Jian He~$^a$~,~~~~~
     Yu-Ping Kuang~$^b$~,~~~~~ C.--P. Yuan~$^c$~} \\[0.3cm]

$^a$~Department of Physics and Institute of High Energy Physics\\
Virginia Polytechnic Institute and State University \\
Blacksburg, Virginia 24061-0435, USA\\
$^b$ China Center of Advanced Science and Technology (~World Laboratory~)\\
     P.O.Box 8730, Beijing 100080, China \\
and Institute of Modern Physics, Tsinghua University, Beijing 100084
China \footnote{Mailing address.}\\
$^c$ Department of Physics and Astronomy, Michigan State University \\
East Lansing, Michigan 48824, USA\\

\end{center}

\bigskip
\centerline{\bf Abstract}
\bigskip

We examine the Lorentz non-invariance ambiguity in longitudinal
weak-boson scatterings
and the precise conditions for the validity
of the Equivalence Theorem (ET).
{\it Safe} Lorentz frames for applying the ET are defined,
and the intrinsic connection between the longitudinal
weak-boson scatterings and probing the symmetry breaking sector is analyzed.
A universal precise formulation of the ET is presented for both the
Standard Model and the Chiral Lagrangian formulated
Electro-Weak Theories. It is shown that in electroweak theories
with strongly interacting symmetry breaking sector,
the longitudinal weak-boson scattering amplitude {\it under proper conditions}
can be replaced by the corresponding Goldstone-boson scattering amplitude
in which all the internal weak-boson lines and fermion loops are ignored.

\noindent
PACS numbers: 11.30.Qc, ~12.15.Ji

\noindent
\begin{center}
--- Accepted for Publication by Physical Review {\bf D}(1995) ---
\end{center}

\newpage
\noindent

\noindent
{\bf 1. Introduction}

The electroweak gauge symmetry is spontaneously broken. As a
consequence of absorbing the corresponding spin-$0$ would-be Goldstone bosons
(GB), the spin-$1$ weak-bosons acquire masses and their longitudinal
components $~V_L^a~$ (~$= W^{\pm}_L, Z^0_L$~) become physical degrees of
freedom. While the transverse components
$~V_T^a~$ (~$= W^{\pm}_T, Z^0_T$~) are irrelevant to the
symmetry breaking (SB) mechanism,
the interactions of the longitudinal weak-bosons (~$V_L^a$'s~)
are expected to be sensitive to probing the SB sector.

Technically, the electroweak Equivalence Theorem (ET) is used to give a
quantitative relation between the $~V_L$-amplitude and the corresponding
GB-amplitude in the high energy region \hbox{(~$~E \gg M_W~$~)},
as shown in Refs.\cite{cl}-\cite{do}.
The most rigorous relation between these two amplitudes (~including all
the possible multiplicative and additive factors~) is given by a general
identity, eq.(1) or (2) in this paper, derived at the level of the LSZ reduced
$S$-matrix elements \cite{hklb}.\footnote{ Similar identities without
the multiplicative factor $~C^a_{mod}$ were given in the early literature
\cite{cl}-\cite{cg}. The appearance of the factor $~C^a_{mod}$ has been
revealed in Refs.\cite{yy}-\cite{do}.
Here we shall adopt the form of the identity generally derived in
Ref.\cite{hklb}.
Other related forms may be found in Refs.\cite{yy}-\cite{do}.}
Based upon this identity we derive the precise
formulation of the ET which is given in this paper as the ensemble of
equations (10) and (10a,b).
By this formulation we show that the ET is not just a technical tool
in calculating a $~V_L$-amplitude using a GB-amplitude, it has an even more
profound physical content for being able to
discriminate processes which are insensitive to probing
the electroweak SB sector.\footnote{To our knowledge,
this point of view has not been given
in the previous literature \cite{cl}-\cite{do}.}

We know that the physical $~V_L$-amplitude can be measured by
experiments and the GB-amplitude, though not directly measurable,
carries information about the SB sector.
Hence, physically, the ET as a bridge
tells us how the $~V_L$-scattering experiments probe the SB sector;
while technically, it replaces the calculation of the $V_L$-amplitude
by a much simpler calculation of the scalar
GB-amplitude in certain energy regime where their difference
can be safely ignored.\footnote{This is an essential simplification
since the $V_L$-amplitude is even much more involved than the
$V_T$-amplitude due to the non-trivial cancellations of large $E$-power
factors from the longitudinal polarization vectors in the high energy region.
This fact was first revealed by Chanowitz and Gaillard \cite{cg}.}
The formulation of the ET in the
Standard Model (SM) and in the Chiral Lagrangian formulated Electro-Weak
Theories (CLEWT) have been recently given in Refs.\cite{hkla}-\cite{hklc},
where the quantization effects and problems related to
the renormalization-scheme and the gauge-parameter dependence
have been systematically studied.\footnote{
A study on the ET in the CLEWT using a non-linear
gauge quantization procedure was recently done in Ref.\cite{do}.}

There is, however, another important problem in this
subject which has not yet been carefully examined. It is about
the Lorentz non-invariance ambiguity in the $~V_L$-scattering amplitudes.
We noticed  that the \hbox{spin-$0$} GB's ( and thus the GB-amplitudes )
are invariant under the proper Lorentz transformation, but both
the longitudinal and the transverse components of the {\it spin-$1$ massive}
weak-bosons ( and thus their scattering amplitudes ) are Lorentz non-invariant
(LNI). After a Lorentz transformation, the longitudinal
 component may mix with the transverse components, and hence the original
$~V_L$-amplitude will become a mixture of
longitudinal and transverse amplitudes. Undoubtedly, one can
 even Lorentz-transform a longitudinal component into a
pure transverse one.\footnote{This can be done by, for example, first boosting
$~V_L~$ to its rest frame and then boosting it in a direction perpendicular to
the first boost.}
Thus a conceptual and fundamental question arises: How can
we use the LNI $~V_L$-amplitudes to probe the
electroweak SB sector of which the physical mechanism should clearly be
independent of the choices of the Lorentz frames?
In this paper, starting from a careful examination of this problem,
we construct a universal precise formulation of the ET which shows that
the $V_L$-amplitudes can probe the electroweak SB sector unambiguously
as long as certain general conditions, as in eqs.(10a,b), are satisfied.

Generally speaking, the {\it replacement} between the $~V_L$-amplitude and the
GB-amplitude (~with possible multiplicative factors~) is LNI unless the
LNI-part in the $~V_L$-amplitude can be ignored.
This LNI-part has the {\it same origin}
as the transverse amplitudes because they can mix or turn into each other under
Lorentz transformations. Hence, the physically important and interesting object
 is the {\it Lorentz invariant {\rm (LI)} part of the $~V_L$-amplitude}.
When we use the GB-amplitude to
predict the physical $~V_L$-amplitude measured by experiments, it does not
distinguish the difference between the experimental results from different
Lorentz frames. Thus, by estimating the LNI-part in the
$~V_L$-amplitude we can determine the accuracy and the validity
region of our quantitative predictions for the physical $~V_L$-amplitudes
based on the ET. We emphasize that the content of the
precise formulation of the ET is
more than just a technical tool for simplifying the  calculations of the
$~V_L$-amplitude. The importance
of the ET is firstly because it provides a conceptual connection
between the would-be Goldstone-boson amplitudes
directly related to the SB mechanism and the experimentally
measurable longitudinal weak-boson amplitudes.
Secondly, as a technical tool,
it may simplify the calculation of the $~V_L$-amplitude which however can
always be directly calculated in spite of its complexity.
Hence the most important task is to find out
the conditions under which the  LNI-part of the $~V_L$-amplitude
can be safely ignored and the LI-part
becomes dominant in the experimentally measured $~V_L$-amplitudes
so that the physical $~V_L$-scatterings can be used to
sensitively  and unambiguously probe the electroweak SB sector.


\vspace{0.4cm}
\noindent
{\bf 2. Avoidance of Lorentz non-invariance ambiguity and the universal
        precise formulation of the ET}

Let us start from the Ward-Takahashi identity
derived in Refs.\cite{cg}-\cite{hklb}:\\
{}~~$<0\,|F^{a_1}_0(k_1)\,F^{a_2}_0(k_2)\cdots F^{a_n}_0(k_n)\,
\Phi_{\alpha}|\,0>\,=\,0~,
{}~~$ in which $~F_0^{a}~$ is the bare gauge fixing function
and $~\Phi_{\alpha}~$ denotes
other possible physical in/out states.\footnote{The
subscript $~\alpha~$ denotes possible Lorentz indices.}
After a rigorous LSZ reduction for the
external $~F^a$-lines, we derived  in Ref.\cite{hklb} the following
general identity for the  renormalized $S$-matrix elements:
$$
\begin{array}{l}
T[V^{a_1}_L,\cdots ,V^{a_n}_L;\Phi_{\alpha} ]
= T[\bar{Q}^{a_1},\cdots ,\bar{Q}^{a_n};\Phi_{\alpha} ]\\[0.4cm]
\bar{Q}^a\equiv -iC^{a}_{mod}\pi^{a} + v^{a}~,
{}~~~~v^a \equiv v^{\mu}V^a_{\mu} ~,
{}~~~~v^{\mu}\equiv \epsilon^{\mu}_L-k^\mu /M_W = {\cal O}(M_W/E) ~,
\end{array}
\eqno(1)                                              
$$
where $~\pi^a$'s  are GB fields.\footnote{Here, $\pi^a$-field by definition
has an opposite sign to that in Ref.\cite{hklb}.
Consequently, the coefficient of $~\pi^a$ in
$\bar{Q}^a$ is $~-i~$ instead of $~+i~$.}
( In this paper, we use $W$ to denote either $W^\pm$ or $Z$, and $E$ is the
energy of the $W$-boson, unless specified otherwise. )
The finite constant modification factor
$~C_{mod}^a~$ has been systematically studied in the
literature \cite{yy}-\cite{do}
and is proved to be
renormalization-scheme and gauge-parameter
dependent.\footnote{The $C^a_{mod}$-factor has also been examined
for the $U(1)$ Higgs theory in Refs.\cite{hkla,hklb} and \cite{kilgore}.}
In General, $~C^a_{mod}~$ is not unity and the
difference $~C^a_{mod}-1~$ comes from loop contributions \cite{yy}-\cite{do}.
A convenient renormalization scheme, {\it scheme-II}, was constructed in
Refs.\cite{hkla}-\cite{hklc} so that
 the modification factors $C^{a}_{mod}$ in both the
SM and the CLEWT are exactly unity, and the application
of the ET is greatly simplified.
It has also been shown that $~~C_{mod}^a-1=
{\cal O}((g^2,\lambda)/16\pi^2)~~$ for the SM with
a light Higgs boson \cite{yy}-\cite{hklb},  and
$~~C_{mod}^a-1={\cal O}(g^2/16\pi^2)~~$ for both
the heavy Higgs SM \cite{yy}-\cite{hklb} and the
CLEWT \cite{hklc}, provided that the GB wavefunction renormalization
constant
$~Z_{\pi^a}~$ is subtracted at a scale $~\mu\sim {\cal O}(M_W)~$
and the physical
mass pole of weak-boson propagator is determined by the on-shell scheme.

The identity in (1) can  be re-written as
$$
T[V^{a_1}_L,\cdots ,V^{a_n}_L;\Phi_{\alpha}]
= C\cdot T[-i\pi^{a_1},\cdots ,-i\pi^{a_n};\Phi_{\alpha}]+ B
\eqno(2)                                               
$$
where
$$
\begin{array}{ll}
C & \equiv C^{a_1}_{mod}\cdots C^{a_n}_{mod} ~; \\
B & \equiv B[v,-i\pi;\Phi_{\alpha}] \\
  & \equiv\sum_{l=1}^n (~C^{a_{l+1}}_{mod}\cdots C^{a_n}_{mod}
T[v^{a_1},\cdots ,v^{a_l},-i\pi^{a_{l+1}},\cdots ,-i\pi^{a_n};\Phi_{\alpha}]
+ {\sf permutations~of~} v's ~{\sf and}~ \pi 's~)  ~.
\end{array}
\eqno(2a,b)                                            
$$
Hereafter we shall  use the shorthand notations $~T[V_L;\Phi_{\alpha}]~$ and
$~T[-i\pi;\Phi_{\alpha}]~$ for $~T[V^{a_1}_L,\cdots,V^{a_n}_L;\Phi_\alpha]~$
and
$~T[-i\pi^{a_1},\cdots,-i\pi^{a_n};\Phi_\alpha]~$, respectively.
Under Lorentz transformations, the
amplitude of spin-$0$ scalar particles is invariant .
If $~\Phi_\alpha~$, in (2), contains no field or
only external physical scalar field(s)
and/or photons, then from (2) the Lorentz non-invariant
$~V_L$-amplitude can be decomposed into two parts. The first part is
$~C\cdot T[-i\pi;\Phi_{\alpha}]~$ which is Lorentz invariant (LI),
and the second
part is the $~v_\mu$-suppressed $B$-term which is  Lorentz non-invariant
 (LNI) because of the external spin-$1$ massive vector field(s).
Such a decomposition
clearly shows the essential difference between the $V_L$-amplitude and
the $V_T$-amplitude since the former
contains a Lorentz-invariant GB-amplitude which is the
intrinsic source causing a large $V_L$-amplitude in the case of
strongly coupled SB sector.
We note that only the LI part of the $V_L$-amplitude
is sensitive to probing the SB sector,
while its LNI part contains a
significant {\it Lorentz-frame-dependent} $~B$-term
and therefore can not be sensitive to the electroweak SB mechanism.

Strictly speaking,
when $~\Phi_{\alpha}~$ contains field(s) such as $V_T$'s and fermions, the
GB-amplitude is not exactly LI due to non-trivial Lorentz transformations of
$~\Phi_{\alpha}~$.  The change of the GB-amplitude due to Lorentz
transformations of $~\Phi_{\alpha}~$ may not be small when compared with the
GB-amplitude itself. For instance, if $~\Phi_{\alpha}~$ contains
a $V_T$-field, this change can
be of the same order of magnitude as the GB-amplitude itself because after a
Lorentz transformation the mixed GB-amplitude (~with one external $V_T$
replaced by $V_L$~) is  only suppressed by $~{\cal O}(M_W/E)~$
(~see the 2nd relation in eq.(7)~), and this suppression factor
is largely compensated by the enhancement factor
$~{\cal O}(E/M_W)~$ arising from the  polarization vector of the resulting
$V_L$. For a fermion field in
$~\Phi_{\alpha}~$, it is easy to see that this change is always
${\cal O}(m_f/E)$-suppressed because this change vanishes in the
$~m_f/E\rightarrow 0~$ limit.
(~Here, $m_f$ and $E$ are the mass and energy of the fermion, respectively.~)
Since the basic properties of the physical mechanism of
the electroweak SB sector are clearly independent of Lorentz frames, this LNI
GB-amplitude (~due to the LNI $~\Phi_{\alpha}~$ field(s)~) would be less
sensitive to probing the SB mechanism.\footnote{
One exception is the top-condensate SM \cite{nambu}
in which the top quark Yukawa coupling is related to
the Higgs boson self-couplings. For $~m_t \approx O(M_W)$, this model
must predict a light Higgs boson which can be
detected through processes other than the $V_L$-scatterings.}
In the case of strongly coupled SB sector, the extra $V_T$('s) and/or
fermion field(s) in $~\Phi_{\alpha}~$ make the leading contribution of the
GB-amplitude contain  more pure gauge couplings and/or Yukawa
couplings (~of the SM fermions~)
and lower $E$-power dependence.
Taking the CLEWT as an example, we easily see that only the pure
scalar vertices contain the largest $E$-power dependence, while all other
vertices containing gauge bosons and/or fermions involve less derivatives
and more gauge and/or Yukawa couplings.
Therefore, in each order of perturbative expansion,
the GB-amplitude  containing the extra $V_T$('s) and/or
fermion field(s) in $~\Phi_{\alpha}~$ is at least
${\cal O}(M_W/E)$- or ${\cal O}(m_f/E)$-suppressed
relative to the pure GB-amplitude (~containing no
external $V_T$ and/or fermion fields~).\footnote{The heaviest
known external fermions are (anti-)top quarks.
Thus $~{\cal O}(m_f/E)\leq {\cal O}(m_t/E)\approx {\cal O}(M_W/E)~$.}

Despite that $~\Phi_{\alpha}~$ might contain some LNI contributions,
 it will not cause the longitudinal-transverse ambiguity
in replacing a longitudinal weak-boson line in the $~V_L$-amplitude
by a corresponding Goldstone-boson line in the GB-amplitude
as long as the LNI $B$-term can be safely ignored.
Thus, we have to find the conditions under which the $~B$-term in (2) is
negligibly small compared with the $~C\cdot T[-i\pi;\Phi_{\alpha}]~$ term.
{\it Such conditions can be conveniently found from (2)
by estimating the magnitude of the $~B$-term from the
analysis of the Lorentz transformation of the $~V_L$-amplitude.}
To estimate the $B$-term,
we first examine how the $~V_L$-amplitude transforms under Lorentz
transformations.\footnote{We
thank Lay Nam Chang for enlightening discussions on this point.}
 Without loss of generality,  let us consider
a Lorentz boost with velocity
$~\beta_0$ along the $\hat{x}$-direction (~from $~oxyzt$-frame
to $~o^\prime x^\prime y^\prime z^\prime t^\prime$-frame~) for an external
longitudinal boson $~V_L~$ (~and also an external transverse boson~)
with momentum $~~ k^\mu = (E,0,0,k) ~~$
in $~oxyzt$-frame:\footnote{ Equivalently, one can study the Lorentz
transformation relation of the spin-1 helicity amplitudes by using the
spin-rotation matrices as shown in Ref.\cite{jw}.
But, here for the purpose of order of magnitude estimate, it is more
convenient to study the Lorentz transformations of the longitudinal
polarization vector $~\epsilon_L^{\mu}~$ and the transverse polarization
vector $~\epsilon_T^{\mu}~$.}

$$
\begin{array}{ll}
 in~~oxyzt~~ frame~: &~~~in ~~o^\prime x^\prime y^\prime z^\prime t^\prime
{}~~ frame:~~\\
k^\mu = (E,0,0,k) ~, &~~~  {k^\prime}^\mu
= (\gamma_0E,-\beta_0\gamma_0E,0,k) ~,\\
\epsilon_L^\mu = \frac{1}{M_W}(k,0,0,E) ~, &~~~
{\epsilon_{(L)}^{\prime}} ^\mu
=\frac{1}{M_W}(\gamma_0k,-\beta_0\gamma_0k,0, E)~,\\
\epsilon_{T_1}^\mu = (0,1,0,0) ~, &~~~ {\epsilon_{(T_1)}^{\prime}}^\mu
=(-\gamma_0\beta_0,\gamma_0,0,0) ~,\\
\epsilon_{T_2}^\mu = (0,0,1,0) ~, &~~~
{\epsilon_{(T_2)}^{\prime}} ^\mu = (0,0,1,0)~.
\end{array}
\eqno(3)                                                     
$$
The three new polarization vectors
in $~o^\prime x^\prime y^\prime z^\prime t^\prime$-frame are  defined as
$$
{\epsilon_L^\prime}^\mu
               =\frac{1}{M_W}(a,-\beta_0\gamma_0^2E^2/a,0,\gamma_0Ek/a)~,~~~~
{\epsilon_{T_1}^\prime} ^{\mu} =  (0,k/a,0,\beta_0\gamma_0E/a) ~,~~~~
{\epsilon_{T_2}^\prime} ^{\mu} =  (0,0,1,0) ~,
\eqno(4)                                                      
$$
where $~~a\equiv \sqrt{ (k^2 +\beta_0^2\gamma_0^2 E^2) }~$,
$\gamma_0 = 1 / \sqrt{1-\beta_0^2}~$,
and $~k' \cdot \epsilon'_\lambda = 0$, for $\lambda=L,\, T_1,\,T_2~$.
 After a little algebra, we get
$$
\begin{array}{l}
{\epsilon_{(L)}^\prime}^\mu -{\epsilon_{L}^{\prime}}^\mu
= b_L{\epsilon_L^\prime}^\mu +\sum_{j=1}^2b_{T_j}{\epsilon_{T_j}^\prime}^\mu~,
{}~~~~{\epsilon_{(T_i)}^\prime}^\mu -{\epsilon_{T_i}^{\prime}}^\mu
= \sum_{j=1}^2h_{i,T_j}{\epsilon_{T_j}^\prime}^\mu +
h_{i,L}{\epsilon_L^\prime}^\mu ~,\\[0.4cm]
b_L =\gamma_0k/a -1 ~,~~
   b_{T_1}= \beta_0\gamma_0M_W/a ~,~~
   b_{T_2}=0~,~~\\
h_{1,T_1}=\gamma_0k/a-1~,~ h_{1,T_2}=0~,~ h_{i,L}=-\beta_0\gamma_0M_W/a~,~~
h_{2,T_j}=h_{2,L}=0 ~.
\end{array}
\eqno(5)                                                 
$$
Hence, for high energy scattering $~~ E\sim k  \gg  M_W ~$,~ we generally have
$$
b_L \approx {\cal O}(M_W^2/E^2) ~,~~~b_{T_j} \leq {\cal O}(M_W/E) ~~;~~~~
h_{i,T_j} \leq {\cal O}(M_W^2/E^2) ~,~~~h_{i,L} \approx {\cal O}(M_W/E) ~~,\\
\eqno(6)                                              
$$

\noindent
where we have taken $~\gamma_0 \geq {\cal O}(1)~$.
Thus, for a boosted external weak-boson field,

$$
\begin{array}{l}
{V^a_{(L)}}^\prime = {\epsilon^{\mu}_{(L)}}^\prime  {V^a_{\mu}}^\prime
\approx [1+{\cal O}(\frac{M_W^2}{E^2})] \, {V_L^a}^\prime
     + \sum_{j=1}^2 {\cal O}(\frac{M_W}{E}) \, {V_{T_j}^a}^\prime~~,\\[0.4cm]
{V_{(T_i)}^a}^\prime = {\epsilon^{\mu}_{(T_i)}}^\prime  {V^a_{\mu}}^\prime
\approx \sum_{j=1}^2 [ 1+ {\cal O}(\frac{M_W^2}{E^2})] \, {V_{T_j}^a}^\prime
                       + {\cal O}(\frac{M_W}{E}) \, {V_L^a}^\prime  ~~.
                                                                \\[0.4cm]
\end{array}
\eqno(7)                                                 
$$

Now, consider the variation
$~\Delta B\equiv B[(v)^\prime,-i\pi;\Phi^{\prime}_{(\alpha)}]
-B[v^\prime,-i\pi;\Phi^{\prime}_{\alpha}]~$, which is the difference between
the boosted amplitude $~B[(v)^\prime,-i\pi;\Phi^{\prime}_{(\alpha)}]~$
and the corresponding amplitude $~B[v^\prime,-i\pi;\Phi^{\prime}_{\alpha}]~$
defined in the  $~o^\prime x^\prime y^\prime z^\prime t^\prime$-frame.
Since the
LNI $B$-term does not contain LI spin-$0$ scalar sub-set which is the only
intrinsic source  that may cause the $~V_L$-amplitude to be large,
the variation
$~\Delta B~$ should be of the same order of magnitude as $B$-term itself, i.e.
$$
{\cal O}(\Delta B)\approx {\cal O}(B[(v)^{\prime},-i\pi ;
\Phi_{(\alpha)}^\prime ])
\approx {\cal O}(B[v^{\prime},-i\pi ;\Phi_{\alpha}^\prime ])
\approx {\cal O}(B[v,-i\pi ;\Phi_{\alpha} ])~~.
$$
Thus we can estimate $~B~$ by
estimating $~\Delta B~$. From (2) and (7),
$$
\begin{array}{ll}
\Delta B
\equiv
B[(v)^{\prime},-i\pi ;\Phi_{(\alpha)}^\prime ]-
B[v^{\prime},-i\pi ;\Phi_{\alpha}^\prime ] =
T[V_{(L)}^{\prime};\Phi_{(\alpha)}^{\prime}] -
T[V_{L}^{\prime};\Phi_{\alpha}^{\prime}] -
C \cdot T[-i\pi;\Phi_{(\alpha)}^\prime - \Phi_{\alpha}^\prime ]\\
\equiv
T[V_{L}^{\prime}+\Delta V_{L}^{\prime};\Phi_{\alpha}^{\prime}
+\Delta\Phi_{\alpha}^{\prime}] -
T[V_{L}^{\prime};\Phi_{\alpha}^{\prime}] -
C \cdot T[-i\pi;\Delta\Phi_{\alpha}^\prime ]\\
=T[\Delta V_{L}^{\prime};\Phi_{\alpha}^{\prime}]+
(~ T[\Delta V_{L}^{\prime};\Delta\Phi_{\alpha}^{\prime}] +
B[v^{\prime},-i\pi ;\Delta\Phi_{\alpha}^{\prime}] ~) ~~~~(~cf.~ (2)~)\\
\approx {\cal O}(T[\Delta V_{L}^{\prime};\Phi_{\alpha}^{\prime}])\\
\approx {\cal O}(\frac{M_W^2}{E_j^2}) \,
  T[ {V_L^\prime} ^{a_1},\cdots , {V_L^\prime} ^{a_n}; \Phi_{\alpha}^\prime ] +
 {\cal O}(\frac{M_W}{E_j}) \, T[ {V_{T_j}^\prime} ^{a_{r_1}},
{V_L^\prime} ^{a_{r_2}},
\cdots , {V_L^\prime} ^{a_{r_n}};\Phi_{\alpha}^\prime ]~~~~~~(~cf.~ (7)~) \\
 \approx {\cal O}(\frac{M_W^2}{E_j^2})\, C \cdot T[ -i\pi^{a_1},\cdots ,
  -i\pi^{a_n}; \Phi_{\alpha}] +
  {\cal O}(\frac{M_W}{E_j})\,C^\prime \cdot
   T[ V_{T_j}^{a_{r_1}}, -i\pi^{a_{r_2}}
      \cdots , -i\pi^{a_{r_n}}; \Phi_{\alpha}] ~~~~~~(~cf.~ (2)~) \\
C= C^{a_1}_{mod}\cdots C^{a_n}_{mod} ~~,~~~~~
C^\prime = C^{a_{r_2}}_{mod}\cdots C^{a_{r_n}}_{mod} ~~.
\end{array}
\eqno(8)                                               
$$
\noindent
Here, in estimating  the order of magnitude of $~\Delta B~$,
we have ignored
$~~T[\Delta V_{L}^{\prime};\Delta\Phi_{\alpha}^{\prime}]~~$ and
$~~B[v^{\prime},-i\pi,\Delta\Phi_{\alpha}^{\prime}]~~$, which
vanish when $~\Phi_{\alpha}~$ contains no field or only scalar(s)
and/or photon(s), and can be at most of the same order
of magnitude as $B$-term itself.
For the same reason, we
have also neglected the LNI-parts  generated from
replacing $~{V_{T_j}^\prime}^{a_{r_1}}~$ and  $~\Phi_{\alpha}^\prime ~$
by $~{V_{T_j}}^{a_{r_1}}~$ and $~\Phi_{\alpha}~$
in the last step of (8). Let $~E_j~$ be the energy
of the $j$-th external longitudinal weak-boson.
 We can thus estimate the order of magnitude
of $~B~$ from (8) by making the $~M_W/E_j$-expansion when
$~~ E_j\sim k_j  \gg  M_W ~~$. Then,\footnote{
As we know, this is the first time that the order of magnitude of the
$B$-term is explicitly given in a general form.}
$$
\begin{array}{ll}
B &= \sum_{l=1}^n (~C^{a_{l+1}}_{mod}\cdots C^{a_n}_{mod}
T[v^{a_1},\cdots ,v^{a_l},-i\pi^{a_{l+1}},\cdots ,-i\pi^{a_n};\Phi_{\alpha}]
+ {\sf permutations~of~} v's ~{\sf and}~ \pi 's~) \\[0.5cm]
& \approx {\cal O}(\frac{M_W^2}{E_j^2})~C \cdot
          T[ -i\pi^{a_1},\cdots , -i\pi^{a_n}; \Phi_{\alpha}] +
  {\cal O}(\frac{M_W}{E_j})~C^\prime \cdot
          T[ V_{T_j} ^{a_{r_1}}, -i\pi^{a_{r_2}}
                      \cdots , -i\pi^{a_{r_n}}; \Phi_{\alpha}]~~. \\
\end{array}
\eqno(9)                                                   
$$
We emphasize that the condition $~~~~ E_j \sim k_j  \gg  M_W , ~~~
(j=1,2,\cdots ,n) ~~~~$ {\it for each external longitudinal weak-boson} is
{\it necessary} in making the $~M_W/E_j$-expansion and ensuring the $~B$-term
(~and its Lorentz variation~) to be
much smaller than$~C\cdot T[-i\pi;\Phi_{\alpha}]~$, as shown in (2).
If the energy of one of
$~V_L^{a_j}$'s is low, say $~~ E_j \sim k_j\approx {\cal O}(M_W) ~~$,
then a Lorentz
transformation may cause large variations in the $V_L$-amplitude and the
Lorentz-frame-dependent  $~B$-term can be as large as
$~C\cdot T[-i\pi;\Phi_{\alpha}]~$, even in the cases where the total energy of
the scattering has already been much larger than $M_W$.

In conclusion, we give following general and  precise
formulation of the ET is \footnote{Here we
still generally keep the modification $~C$-factor in the ET.
The exact simplification of the $~C$-factor
as unity has been given before for both the SM \cite{hkla,hklb}
and the CLEWT \cite{hklc}.}
$$
T[V^{a_1}_L,\cdots ,V^{a_n}_L;\Phi_{\alpha}]
= C\cdot T[-i\pi^{a_1},\cdots ,-i\pi^{a_n};\Phi_{\alpha}]+
{\cal O}(M_W/E_j{\sf -suppressed} ),
\eqno(10)                                               
$$
and, from eqs.(2b) and (9), the conditions for ignoring the LNI
and $~M_W/E_j$-suppressed $B$-term
on the RHS of (10) are:
$$
\begin{array}{l}
E_j \sim k_j  \gg  M_W , ~~~~~(~ j=1,2,\cdots ,n ~)~~;\\
B  \ll  C\cdot T[-i\pi^{a_1},\cdots ,-i\pi^{a_n};\Phi_{\alpha}] ~~.\\
\end{array}
\eqno(10a,b)                                   
$$

Before going to detailed discussions, we first point out several important
features contained in the above formulation.
Firstly, the second term on the RHS of (10),
i.e. the $B$-term, as emphasized is
only $~~{\cal O}(M_W/E_j{\sf -suppressed} )~~$ relative to the
leading contributions in $~C\cdot T[-i\pi;\Phi_{\alpha}]~$, and therefore
is {\it not necessarily} of the $~~{\cal O}(M_W/E_j)~~$ in magnitude.
As clearly shown in (9), the magnitude of the $B$-term explicitly
depends on the size of the amplitudes
$~~T[ -i\pi^{a_1},\cdots , -i\pi^{a_n}; \Phi_{\alpha}]~~$ and
$~~T[ V_{T_j} ^{a_{r_1}}, -i\pi^{a_{r_2}}
   \cdots , -i\pi^{a_{r_n}}; \Phi_{\alpha}]$.
Consequently, {\it the $B$-term itself can be either larger or smaller than
$~{\cal O}(M_W/E_j)~$.}
For example, as we shall prove in the following, the largest $B$-term
in the CLEWT is of  $~{\cal O}(g^2)~$, cf. eq.(17).
Secondly, the actual suppression
factor in the $B$-term is $~M_W/E_j~$ instead of
$~M_W/\sqrt{s}~$ as appeared in some current literature. (~$\sqrt{s}$
 is the total center-of-mass energy of  the scattering.~)
So, condition (10a) is usually stronger than $~\sqrt{s}\gg M_W~$.
The existence of the condition (10b) for the CLEWT has been recently
pointed out in Refs.\cite{hklc}-\cite{do}. Here we emphasize that
(10b) generally exists for {\it any perturbation expansion},
not only for the chiral perturbation expansion,
but also for the usual loop-expansion ( adopted in the SM )
and the large ${\cal N}$ expansion,
etc.\footnote{This general fact, as we know, has not been revealed before.}
This will be examined in detail later.
Thirdly, the {\it equivalence} theorem is about the ``equivalence''
between the $V_L$-amplitude and the GB-amplitude (~not the GB-amplitude
plus the $B$-term~). Therefore it is important to give explicit conditions,
i.e. (10a) and (10b), under which
the $~M_W/E_j$-suppressed $B$-term in (10) can be ignored to establish
the equivalence between the $V_L$-amplitude and the GB-amplitude.
It is clear that one can technically improve the prediction of the
$V_L$-amplitude from the RHS of (10) by including
the complicated $B$-term ( or part of
$B$ ) \cite{ger}, but this is not an improvement of
the equivalence between the $V_L$- and the GB-amplitudes.
As noted in our above discussion,
the LNI $B$-term has the same origin as the transverse amplitudes
and is thus insensitive to probing the electroweak SB sector.
More specifically, even for the CLEWT with strongly coupled SB sector,
the largest $B$-term is of $~{\cal O}(g^2)~$ (cf. eq.(17) or (21)),
which depends only on the electroweak gauge coupling and is not sensitive to
the interactions responsible for the electroweak symmetry breaking.
( The same conclusion holds for the leading amplitudes of
pure transverse gauge boson scatterings. )
Therefore, for the longitudinal weak-boson scattering processes to be
sensitive to the electroweak SB sector, the conditions
(10a) and (10b) must be satisfied such that
the scalar GB-amplitudes can {\it dominate} the
contributions to the $V_L$-amplitudes.

Let us further analyze the important implications of eqs.(10a) and (10b)
in details.  First, we note that the condition (10a)  defines the
{\it safe} Lorentz frames for the precise formulation and the
application of the ET.
As we  pointed out, a longitudinal weak-boson can
turn into a mixture of longitudinal and transverse state under Lorentz
transformations while the scalar Goldstone boson is invariant.
This implies that (10) cannot hold in all
Lorentz frames. To resolve this
longitudinal-transverse ambiguity, a set of {\it safe} Lorentz frame
has to be defined such that
for {\it each} external $V_L$ particle
$~E_j \gg M_W~$.\footnote{Here we do not take
the {\it unphysical limit} as
$~~M_W ( = g f_\pi /2 ) \rightarrow 0~~$, which requires either
the gauge coupling $~g=0~$,
implying no Higgs mechanism and the disappearance of physical
longitudinal component of the $W$-boson, or the vacuum expectation value
$~f_\pi =0~$, in contradiction with
the non-vanishing physical Fermi-scale and
the presence of the electroweak symmetry breaking. Such limits are actually
{\it unnecessary} for the precise formulation of the ET.}
This means that {\it $V_L$ is sensitive to probing the {\rm SB} sector
only in the sufficiently high energy region where the $V_L$, originally
coming from ``eating'' the {\rm GB},
mainly behaves like the {\rm GB}, and the effects of its mixing with
the transverse components are always
$M_W/E_j$- or $(M_W/E_j)^2$-suppressed and negligibly small.}
If we change this high energy property by making
Lorentz transformations such that $M_W/E_j\approx {\cal O}(1)$, this
longitudinal-transverse ambiguity can no longer be ignored and
the LNI-part of $~T[V_L;\Phi_{\alpha}]~$
will be of the same order of magnitude as
the LI-part of $~T[-i\pi;\Phi_{\alpha}]~$ (cf. (9)).

The condition (10a) is actually
quite strong. Naively one may expect that requiring the total
center-of-mass (CM) energy $~~ E_{\rm CM} \gg M_W~~$  can
always guarantee the equivalence of the $~V_L$-amplitude and
the GB-amplitude. However, we shall show as follows
 that even in the SM, there are counter examples to this weaker condition
in which only $~~ E_{\rm CM} \gg M_W~~$ is satisfied but the (10a) is violated.
Subsequently,  eq.(10) does not hold.
To illustrate this point, we consider the scattering
process $~Z_L+H\rightarrow Z_L+H~$,
where $~H~$ is the SM Higgs particle. In the CM frame of $Z_L H$,
the exact tree-level $~Z_L$- and GB-amplitudes are:
$$
\begin{array}{ll}
T[Z_LH \rightarrow Z_LH] & =  \displaystyle
 ig^2\left[ \frac{p^2(1-\cos\theta)-M^2_Z\cos\theta}
{2M^2_Z}~\frac{t +2m^2_H}{t-m^2_H}+[p^2(1-\cos\theta)-M^2_Z\cos\theta]\cdot
\right.\\[0.4cm]
& ~~~~\displaystyle\left. \left(\frac{1}{u-M^2_Z}+\frac{1}{s-M^2_Z}\right)
-\frac{p^2}{M^2_Z}\left(\frac{(\cos\theta\sqrt{p^2+M^2_Z}+
\sqrt{p^2+m^2_H})^2}{u-M^2_Z}+
\frac{s}{s-M^2_Z}\right) \right] ~~,\\[0.6cm]
T[\pi^0 H \rightarrow \pi^0 H] & = \displaystyle
i\left[ -\frac{m^2_H}{f_\pi^2}\frac{t+2m^2_H}{t-m^2_H}-\frac{m^4_H}{f_\pi^2}
\left(\frac{1}{u-M^2_Z}+\frac{1}{s-M^2_Z}\right)+\frac{g^2}{4}\left(\frac{s-t}
{u-M^2_Z}+\frac{u-t}{s-M^2_Z}\right)\right] ~~,\\
\end{array}
\eqno(11)                                
$$

\noindent
where $~p~$ is the CM momentum, $~\theta~$ is the
scattering angle and $~s,~t,~u~$ are the Mandelstam variables.
We consider two typical high energy limits: $~E_{\rm CM} \gg m_H\sim M_Z~$
and   $~E_{\rm CM} > m_H \gg M_Z~$,
where $~E_{\rm CM}=\sqrt{s}~$ is the total energy.
In the first case, the energy of the $Z$-boson
$~~E_Z\sim p \gg M_Z~~$ so that our new
condition (10a) is satisfied; while in the second case, $~~E_Z\sim p
\sim {\cal O}(M_Z)~~$ which violates the (10a).
In both cases the conventional condition $~~ E_{\rm CM} \gg M_Z~~$
is satisfied.
\\
(i). For the first case $~E_{\rm CM} \gg m_H\sim M_Z~$,
which implies $~~E_Z\sim p \gg M_Z~$,
(11) gives
$$
\begin{array}{ll}
T[Z_L H\rightarrow Z_L H] &
=\displaystyle -i\left[ \frac{m^2_H}{f_{\pi}^2}+\frac{g^2}{4}
\frac{3+\cos^2\theta}{1+\cos\theta}\right]
+{\cal O}(g^2M^2_Z/p^2,\lambda m^2_H/p^2) ~~,\\[0.3cm]
T[\pi^0 H\rightarrow \pi^0 H] &
=\displaystyle -i\left[ \frac{m^2_H}{f_{\pi}^2}+\frac{g^2}{4}
\frac{3+\cos^2\theta}{1+\cos\theta}\right]+{\cal O}(g^2M^2_Z/p^2,
\lambda m^2_H/p^2)~~,\\[0.5cm]
T[Z_LH\rightarrow Z_L H] &
=T[i\pi^0 H\rightarrow -i\pi^0 H] + {\cal O}(g^2M^2_Z/p^2,\lambda m^2_H/p^2)
{}~~.\\
\end{array}
\eqno(12)                                  
$$
Thus, the $~V_L$-amplitude is equivalent to the GB-amplitude,
and can be used to probe the SB sector.
In this case, the CM frame is a {\it safe} frame in applying the ET.
\\
(ii). For the second case $~E_{\rm CM} > m_H \gg M_Z~$,\footnote{
For example, $~E_{\rm CM}=1\,$TeV,~$m_H=800\,$GeV.}~
which implies $~~E_Z\sim p \sim {\cal O}(M_Z)~$,~ (11) gives
$$
\begin{array}{l}
T[Z_LH\rightarrow Z_LH]
= \displaystyle i4\frac{(p^2+M^2_Z)\cos\theta -3p^2}{f_\pi^2}
                         + {\cal O}(p/m_H, M_Z/m_H) ~~,\\[0.4cm]
T[\pi^0 H\rightarrow \pi^0 H]~~~
= \displaystyle i2\frac{-2p^2(1-\cos\theta )+M^2_Z}{f_\pi^2}
                         + {\cal O}(p/m_H, M_Z/m_H) ~~,\\[0.5cm]
T[Z_LH\rightarrow Z_LH]-T[i\pi^0 H\rightarrow -i\pi^0 H]
=\displaystyle i2\frac{-4p^2+M^2_Z(2\cos\theta -1)}{f_\pi^2}
               + {\cal O}(p/m_H,M_Z/m_H) ~.\\
\end{array}
\eqno(13)                                          
$$
As shown in the above equation, the difference between the $~V_L$-amplitude
and the GB-amplitude has the same size as the $~V_L$-amplitude itself.
Thus, the $~V_L$-amplitude is not equivalent to the GB-amplitude.
The CM frame in this case is therefore {\it not a safe} frame
for applying the ET
because in this frame our condition (10a) is violated.

Next, we examine the condition (10b) for ignoring the LNI $~B$-term,
which is the sum of all the $~v_\mu$-suppressed terms in (2).
Based upon the order of magnitude estimate of the $~B$-term given
in eq.(9), we can further express the (10b) as
$$
\begin{array}{l}
{\cal O}(\frac{M_W^2}{E_j^2}) \, T[-i\pi^{a_1},\cdots,-i\pi^{a_n};\Phi_\alpha]
+{\cal O}(\frac{M_W}{E_j}) \, T[ V_{T_j}^{a_{r_1}}, -i\pi^{a_{r_2}},
                      \cdots , -i\pi^{a_{r_n}}; \Phi_{\alpha}]
 \ll  T[-i\pi^{a_1},\cdots ,-i\pi^{a_n};\Phi_{\alpha}] ~.
\end{array}
\eqno(14)                                            
$$
Here we have dropped the factor $~1/C_{mod}^{a_{r_1}}~$ in the
second term on the  LHS since we can always adopt the {\it Scheme-II}~
of  Refs.\cite{hkla}-\cite{hklc} to make $~C_{mod}^a\equiv 1$.
Even in some other schemes as described in the paragraph just below eq.(1),
$~ C_{mod}-1~$ is of $~{\cal O}((g^2,\lambda)/16\pi^2)~$
and  $~{\cal O}(g^2/16\pi^2)~$  for
the light Higgs SM and the heavy Higgs SM ( or the CLEWT ),
respectively, so that $~1/C_{mod}^{a_{r_1}}~$  will not affect the
order of magnitude estimate on the LHS of (14) since only the leading terms
are relevant.
The condition (14) shows that after ignoring the $B$-term,
we only need to keep in the GB-amplitude the contributions
that satisfy the condition in (14).  If we further make a
perturbative expansion on the GB-amplitude, (14) would then
constrain  the smallest term to be included in the GB-amplitude
for a fixed energy, or the lowest energy required to calculate
the GB-amplitude to a desired accuracy.

In perturbative calculations,
we may make loop expansion with the expansion parameter
$~\hbar~$, or the momentum expansion with the expansion
parameter $~E/\Lambda~$, or the large ${\cal N}$
expansion with the expansion parameter $~1/{\cal N}~$, etc.
Practically we can only calculate the amplitude $T$ to a {\it finite} order
in the perturbation expansion, i.e.
$~~~~
T= \sum_{\ell=0}^N T_\ell = \sum_{\ell=0}^N \bar{T}_\ell \alpha^{\ell}
{}~~~$,~
where $~\alpha~$ denotes the expansion parameter.
In perturbative expansion, we have  $~~ T_0 > T_1,~ T_2, \cdots, T_N ~$. Let
$~~ T_{min}~$  be the smallest one in the set
$~~\{ T_0, T_1, \cdots , T_N \}~$.~
The condition (14) then implies\footnote{For special cases with
{\it both} $T_0$-amplitudes on the LHS of (15) vanishing, the non-trivial
condition is given via replacing the two $T_0$-amplitudes by corresponding
higher order amplitudes of maximum values among
$~~ T_1, \cdots , T_N ~$.~ In this case, (15) simply reduces to (10a) up to
next-to-leading order. Explicit examples of such kind are discussed
in detail elsewhere.}
$$
\begin{array}{l}
{\cal O}(\frac{M_W^2}{E_j^2}) \,T_0[ -i\pi^{a_1},\cdots , -i\pi^{a_n};
\Phi_{\alpha}] +
  {\cal O}(\frac{M_W}{E_j}) \,T_0[ V_{T_j}^{a_{r_1}}, -i\pi^{a_{r_2}},
                      \cdots , -i\pi^{a_{r_n}}; \Phi_{\alpha}]
 \ll  T_{min}[-i\pi^{a_1},\cdots ,-i\pi^{a_n};\Phi_{\alpha}] .
\end{array}
\eqno(15)                                                    
$$
When $~N=0~$, i.e. only the leading order in the expansion is kept,
(15) reduces to (10a).
Hence, {\it to leading order in any perturbative expansion, the condition
{\rm (10a)}
 is always sufficient to ensure the smallness of the $~B$-term.}
The extra condition (15) is non-trivial only if higher order contributions are
included.\footnote{For example, in the $~1/{\cal N}$-expansion formalism,
some previous studies \cite{ey} applied the ET only to leading
order so that condition (15) is unnecessary there.
The specific form of (15) in the $~1/{\cal N}$-expansion beyond leading order
will be given elsewhere.} This is why in many previous tree-level calculations
for the $V_L$-amplitudes the ET was found to work well after the condition
(10a)
is satisfied.
Actually, when applying the ET to any perturbation theory,
two kinds of expansions have to be considered:
one is the expansion in $\alpha$, the intrinsic expansion parameter
of the theory itself;  another is the expansion in power of
$~M_W/E_j$, as required by the ET (~cf. eq.(10)~).
In the first expansion we usually try to include contributions
beyond the leading order, while in the second expansion
we always keep only  the leading order term for both the physical
and the technical reasons  explained above. The
condition (15) is required to ensure the $~M_W/E_j$-suppressed~
$B$-terms from the leading order in $\alpha$ to be much
smaller than the smallest term $T_{min}[-i \pi ;\Phi_\alpha]$
kept in the GB-amplitude.
If (15) is satisfied, i.e. (10b) is satisfied,  the $~V_L$-amplitude is
equivalent to the GB-amplitude.  Thus in this case, the $~V_L$-amplitude
 can be given by a much simpler calculation of
 the GB-amplitude. This is the technical aspect of (10).
Physically, {\it the applicability of (10) implies that this
$~V_L$-amplitude is sensitive to probing the SB sector to the accuracy
of $T_{min}[-i \pi ;\Phi_\alpha]$.}
If (15) is not satisfied, i.e. the smallest term kept in the
GB-amplitude does not dominate the
LNI and  $~M_W/E_j$-suppressed~  $B$-term,
 then (10b) is not satisfied, therefore (10) is not true.
Hence, the $~V_L$-amplitude and the GB-amplitude are not equivalent,
and this $~V_L$-scattering
process cannot be sensitive to probing the electroweak SB sector
to the  accuracy of $T_{min}[-i \pi ;\Phi_\alpha]$.
In addition to its {\it technique content} as a tool in simplifying
the $~V_L$-amplitude calculations, the above formulation
of the ET, eqs.(10) and (10a,b), {\it has a
profound physical content in discriminating processes which are
insensitive to probing the electroweak
SB sector to certain required precision.}

To illustrate the condition (15), we consider two
typical examples with $~N=1~$, i.e. up to
the next-to-leading order. They are the high
energy $~2\rightarrow 2~$ pure $~V_L$-scatterings
predicted in the
CLEWT, and in the SM with a light Higgs boson (~$~m_H \ll E~$~).
We shall work in the CM frame of $V_L$-$V_L$
which is a safe Lorentz frame for $M_W \ll E$.

First, we examine (15) in the CLEWT, where the SB sector
is non-linearly realized and strongly interacting.
Now $~T_0$ and $~T_1$ are the $~E^2$-level and the $~E^4$-level contributions,
respectively. By a direct power counting \cite{power},
these scattering amplitudes are found to behave as
$$
\begin{array}{l}
T_0 [ \pi^a\pi^b\rightarrow\pi^c\pi^d ]
= {\cal O}\left(\displaystyle\frac{E^2}{f_\pi^2}\right) ~~,~~~~
T_0 [ V_T^a\pi^b\rightarrow\pi^c\pi^d ]
= {\cal O}\left( g\displaystyle\frac{E}{f_\pi}\right) ~~,\\[0.4cm]
T_1 [ \pi^a\pi^b\rightarrow\pi^c\pi^d ]
= {\cal O}\left(\displaystyle\frac{E^2}{f_\pi^2}
            \displaystyle\frac{E^2}{\Lambda^2}\right) ~~,\\[0.5cm]
\end{array}
\eqno(16)                                             
$$
where $~~ \Lambda \simeq 4\pi f_{\pi} \simeq 3\,$TeV
is the cut-off of the CLEWT
according to the usual dimensional analysis \cite{georgi}.
The order of magnitude estimates in (16) are easy to understand.
For the amplitude
$~~T_0 [ \pi^a\pi^b\rightarrow\pi^c\pi^d ] ~~$, it is just the standard
low energy theorem result \cite{low}, where the dimensionful
scale factor in the denominator is $~f_\pi^2~$, not $~\Lambda^2
\simeq (4\pi f_\pi )^2~$.~
The amplitude
$~T_0 [ V_T^a\pi^b\rightarrow\pi^c\pi^d ]~$ with one external transverse
gauge boson can at most be of  $~{\cal O}(g\frac{E}{f_\pi})~$
because any vertex with only one gauge boson line
will contain a factor $g$ and one less  partial derivative than
the corresponding  GB-vertex.
The next-to-leading order amplitude
$~~T_1 [ \pi^a\pi^b\rightarrow\pi^c\pi^d ] ~~$ is well-known to be
$~E^2/\Lambda^2$-supressed relative to the leading order contribution
$~~T_0 [ \pi^a\pi^b\rightarrow\pi^c\pi^d ] ~~$ due to the momentum
expansion in the CLEWT.
Substituting (16) into (15), we find that the largest $B$-term gives
$$~~
B={\cal O}(g^2) ~~,
\eqno(17)                                              
$$
which also coincides with a previous explicit calculation for the
$~W^+_LW^-_L\rightarrow Z_LZ_L~$ scattering \cite{chano}. Thus,
the condition (15) for ignoring the $B$-term in the CLEWT is
$~~~{\cal O}(g^2) \ll \frac{E^2}{f_\pi^2}\frac{E^2}{\Lambda^2}~$.\footnote{
This is different from the condition derived in Ref.\cite{do}, for example,
in which the $B$-term
was estimated as $~{\cal O}(M_W/E)~$ instead of
$~{\cal O}(g^2)~$.
( See the 2nd inequality in the eq.(27) of the 1st paper or the eq.(65) of the
2nd paper in Ref.\cite{do}. ) The authors of Ref.\cite{do} kindly informed us
recently that their new analyses (in preparation) agreed with
our condition (18).}
After replacing $g^2$ by $({2 M_W / f_\pi})^2$, we obtain
$$
\frac{M_W^2}{E^2}  \ll  \frac{1}{4} \frac{E^2}{\Lambda^2} ~~,~~~~{\rm or}~~~~~~
(0.7\,{\rm TeV}/E)^4  \ll  1 ~~.
\eqno(18)                                              
$$
{}From (18), we see that the higher the energy $E$ is, the better the condition
(18) is satisfied.
For examples, for $~E=800\,$GeV, $1\,$TeV, and $1.3\,$TeV, (18) gives
$~~0.56 \ll 1~$,~ $~~0.23 \ll 1~~$ and $~~0.081 \ll 1~$,~ respectively.
These numerical results indicate that the ET technically works well if
$~~E\geq 1\,$TeV.\footnote{When the energy $E$ is close to the effective
cut-off $~\Lambda~$ of the CLEWT, the higher order
corrections in the momentum expansion become important
and should be included, but it does not necessarily
imply a violation of the ET.}
Most importantly, it
also tells us that in order to sensitively
probe the strongly interacting SB sector,
up to the order of $E^4$, we {\it must}
raise the collider energy far beyond the TeV region
so that there will be enough $V_L$-$V_L$ luminosities in the TeV region
for $V_L V_L \rightarrow V_L V_L$ scatterings.
In this example, we assume that there is no light resonance
(~defined as a resonance with mass much less than $1\,$TeV~) involved
in the pure $~V_L$-scattering.
Next, let's examine what if there is a resonance, such as a SM
Higgs boson, far below TeV.

In the case of the SM with $~m_H, M_W \ll E~$,
the one-loop level $~2\rightarrow 2~$
scattering amplitude $~T_1~$ is of the order
\footnote{Since the $U(1)_{em}$ gauge coupling
$e$ is suppressed by $~\sin\theta_W\approx 0.48~$  relative to $~g~$,
it is sufficient to take $~g~$ for the order of magnitude estimate.}
$$
\begin{array}{ll}
T_1 [\pi^{a_1},\cdots ,\pi^{a_4} ] & \approx
\displaystyle
{\cal O}\left(\frac{g^2,\lambda}{16\pi^2}\right)
                ~T_0[\pi^{a_1},\cdots ,\pi^{a_4}]~~,\\[0.3cm]
T_0[ V_{T_j}^{a_{r_1}}, \pi^{a_{r_2}},
                      \cdots , \pi^{a_{r_4}}]
& \displaystyle\approx
{\cal O}\left(\frac{M_W}{E}\right) ~T_0[\pi^{a_1},\cdots ,\pi^{a_4}]~~,\\
\end{array}
\eqno(19a,b)                                       
$$
where the factor $~~1/16\pi^2~~$ (~$=\pi^2 / (2 \pi)^4 $~)
is the  characteristic of
each loop correction.\footnote{(19a) also coincides with previous explicit
1-loop calculations \cite{dw}.}
Thus (15) and (19a,b) give
$$
{\cal O}\left(\frac{M_W^2}{E^2}\right)
      \ll  {\cal O}\left(\frac{g^2,\lambda}{2\cdot 16\pi^2}\right)~~,
{}~~~{\rm or}~~~~~~~ \left(\frac{1.4\,{\rm TeV}}
{{\cal O}(g,\sqrt{\lambda})\cdot E}\right)^2 \ll  1~~,
\eqno(20)                                       
$$

\noindent
which is a rather strong condition. For $~\lambda\approx 10\,g^2~$, i.e.
$~m_H = \sqrt{2 \lambda} \, f_\pi \approx 700\,$GeV,
the condition (20) requires
$~(0.7\,{\rm TeV}/E)^2 \ll 1~$.
For $~E=1\,$TeV, $1.3\,$TeV,  and $2\,$TeV, (20) gives
$~~0.49 \ll 1~$,~ $~~0.29 \ll 1~~$ and $~~0.12 \ll 1~$,~ respectively.
For $~\lambda\approx g^2~$, i.e.
$~m_H\approx 225\,$GeV, (20) means $~~(2.2\,{\rm TeV}/E)^2 \ll 1~~$, which
requires $E$ be at least a few TeV to
probe the SB sector of the SM with a light Higgs
boson to the accuracy of including loop corrections
in the GB-amplitude.
This is however not a disaster because to probe the SB sector of
the SM with
a light Higgs boson we would have to search for a light resonance in the
region $E_{\rm CM} \sim m_H$. It has been extensively studied in the
literature how to detect such a SM Higgs boson resonance
through other production mechanisms other than the
$~V_L$-$V_L$ fusion process
at the LHC (~Large Hadron Collider, ${\rm p} {\rm p}$~), the
NLC (~Next Linear Collider, $e^-e^+$~), and some photon-photon
linear colliders~\cite{hunter,bz}.
Because the $~V_L$-$V_L$ scattering amplitude in the SM is
unitary, if the SM Higgs boson is not heavy,
the $~V_L$-$V_L$ scattering amplitude in the vicinity of $1\,$TeV
can never be large enough to be useful for
 probing the SB sector of the SM with a light Higgs resonance.
Our condition (20) sets the lower limit of the energy range in which the ET
can be used to calculate $~T[V_L;\Phi_{\alpha}]~$ in terms of
$~T[-i\pi;\Phi_{\alpha}]~$ to the
accuracy of including one loop corrections in the SM with  $~m_H\ll E~$.

\vspace{0.4cm}
\noindent
{\bf 3. The ET for pure longitudinal scatterings in probing strongly
        coupled SB sector}

Here we give a further discussion on the precise formulation of the ET for
pure longitudinal weak-boson
scatterings in the case of a strongly interacting SB sector.
We first estimate the largest contribution in the $B$-term, as defined in
(2), based upon the eq.(15) and the results from
a precise power counting \cite{power}.
For both the SM with a heavy Higgs boson,
$~ m_H \gg E~$,
and the general CLEWT, we find that
$~B~$ is of $~{\cal O}(g^2)f_\pi^{D_T}~$, where $~D_T~$
is the dimension of the scattering amplitude $T$, and $~D_T=4-n_{e}~$,
for $n_e$ external $V_L$- or GB-lines. This is only a direct generalization
of our above counting result (17) for the $~2\rightarrow 2~$ scattering
with $~n_e=4~$.
(~For pure longitudinal weak-boson scatterings,
the minimum $g$-dependence in the $B$-term is of
$~{\cal O}(g^2)~$ because based upon the eq.(9) or
the LHS of the eq.(15) the $g$-dependence can arise either from
the factor $~{\cal O}(M_W^2/E_j^2)~$ ( containing a $g^2$-factor ) or
from the factor $~{\cal O}(M_W/E_j)~$ ( containing a $g$-factor ) and
the additional $g$-factor accompanying with each gauge boson
field $~V_T^{a_{r_1}}~$.~)
It is easy to see that
in the GB-amplitudes all tree level Feynman graphs with internal
gauge boson line(s) are at most of
$~{\cal O}(g^2)f_\pi^{D_T}~$, i.e. of the same order
as the largest contribution in $B$, because
 one internal gauge boson line will
induce an extra $g^2$-factor from the two vertices attached to it
and reduce the $E$-power by a factor of $2$ as compared
with the tree-level diagrams with only pure GB-lines
which are of the order $~{\cal O}(\frac{E^2}{f_{\pi}^2})f_{\pi}^{D_T}~$
as given by the low energy theorem \cite{low}.
For higher loops or higher dimensional operators, the
graphs with internal gauge-boson line(s) will
be suppressed by higher powers of $E/\Lambda$.
Thus, beyond the tree level, all graphs in
the GB-amplitudes with internal
gauge boson line(s) are at most of
$~{\cal O}(g^2\frac{E^2}{\Lambda^2})f_\pi^{D_T}~~$.\footnote{
For the SM with a heavy Higgs boson, $\Lambda$ is replaced by $m_H$.
For the CLEWT, $\Lambda$ is taken to be about $4 \pi f_\pi$.}
Therefore, once we ignore the largest $~B$-terms according to
the condition (10b) or (15), we should also correspondingly ignore all the
GB-graphs with internal gauge-boson lines  to all orders in
the heavy Higgs mass expansion or the momentum expansion.
Furthermore, fermion fields can only appear in loops in the
GB-amplitudes, their contributions are at most of
$~{\cal O}(y_f^2\frac{E^2}{\Lambda^2})f_\pi^{D_T}$ \cite{power}, where
$y_f \leq y_t \approx {\cal O}(g)$ and
$y_f$ is the Yukawa coupling of fermion $f$.
(~Here we assume all possible non-SM heavy fermions have been
integrated out in the CLEWT.~)
Thus, their contributions should also be ignored once the
$B$-term, of ${\cal O}(g^2)f_\pi^{D_T}$, is ignored.

In conclusion, for pure longitudinal weak-boson scatterings
in theories with the strongly
interacting SB sector, the ET (~eqs.(10) and (10a,b)~)
 can be further simplified as
$$
\begin{array}{l}
T[V^{a_1}_L,\cdots ,V^{a_n}_L]
= \bar{C}\cdot T[-i\pi^{a_1},\cdots ,-i\pi^{a_n}]|_{g,e,y_f=0}
                                  + {\cal O}(g^2)f_\pi^{D_T}\\[0.3cm]
E_j \sim k_j  \gg  M_W , ~~~~~(~ j=1,2,\cdots ,n ~)~~,\\
{\cal O}(g^2)f_\pi^{D_T}
   \ll  \bar{C}\cdot T[-i\pi^{a_1},\cdots ,-i\pi^{a_n}]|_{g,e,y_f=0} ~~,\\
\bar{C} = \bar{C}_{mod}^{a_1}\cdots \bar{C}_{mod}^{a_n}~~,~~~~~~
\bar{C}_{mod}^{a}=C_{mod}^a|_{g,e,y_f=0}
 ={\left(\displaystyle\frac{M_a}{M_a^{phys}}\sqrt{\frac{Z_{V^a}}{Z_{\pi^a}}
   }Z_{M_a}\right)}\displaystyle\left|_{g,e,y_f=0} \right.  ~,
\end{array}
\eqno(21,21a,b,c)                                
$$
where $~~\pi^a_0=\sqrt{ Z_{\pi^a} } \pi^a~,~V^a_0=\sqrt{ Z_{V^a} } V^a~~$
and $~{M_a}_0=Z_{M_a} M_a ~$.
 $\pi^a_0$ and $V^a_0$ are bare fields, and
$~M_a=M_W$ or $M_Z~$. $~M_a^{phys}~$ denotes the physical mass of the
$W^{\pm}$ or $Z^0$ boson and is equal to $M_a$ only in the on-shell
renormalization scheme \cite{hkla}-\cite{hklc}.
We note that in the above equations, the condition $~g,e,y_f=0~$
is meant to ignore all the gauge coupling or Yukawa coupling
dependent contributions in the GB-amplitudes
after replacing $M_W$ and $M_Z$ (~or $m_f$~) by the products of
$g$ (~or $y_f$~) and $f_\pi$, because they are at most
of the same order as $B$-term.
The $g^2$- and $y_f^2$-dependent
terms in the modification factor (~$C_{mod}^a -1$~)
come from loop corrections and are at
most of ${\cal O}({ g^2,y_f^2 \over 16\pi^2})
\leq {\cal O} ( g^2 {f_\pi^2 \over \Lambda^2})~~$ \cite{yy}-\cite{do}.
(~Recall that $y_f \leq {\cal O}(g)$.~)
This modification factor times the
largest term in the GB-amplitude,
of ${\cal O} ({E^2 \over f_\pi^2})f_\pi^{D_T}$, can only be of
${\cal O}( g^2 {E^2 \over \Lambda^2})f_\pi^{D_T}$, which is again
${E^2 \over \Lambda^2}$-suppressed relative to the $B$-term and
should be ignored. Then we find that those complicated $~\Delta_i$-quantities
inside of $~C_{mod}^a~$, as defined in \cite{hkla}-\cite{hklc}, disappear
after ignoring all $g^2$- and $y_f^2$-dependent terms.
So we can make the finite modification $C$-factor
{\it exactly unity} by
simply choosing the unphysical wavefunction renormalization constant
$~Z_{\pi^a}~$ as
$$
\begin{array}{l}
Z_{\pi^a}
= {\left(\displaystyle\left(\frac{M_a}{M_a^{phys}}\right)^2
    Z_{V^a}\cdot Z_{M_a}^2 \right)} \displaystyle\left|_{g,e,y_f=0}\right.~~,
{}~~~~~~~~~(~Scheme-III~)\\
\bar{C}_{mod}^a =C_{mod}^a|_{g,e,y_f=0} =1 ~~.
\end{array}
\eqno(22)                                         
$$
We call the above
renormalization prescription as~{\it Scheme-III}~ in which all other
renormalization conditions can be freely chosen as in any of the standard
renormalization schemes.

In the general CLEWT,
up to the $~E^4$-level, the pure GB-amplitude without internal gauge boson
lines can be easily counted as of
the form $~{\cal O}(1) f_{\pi}^{D_T}\frac{E^2}
{f_\pi^2}\frac{E^2}{\Lambda^2} ~$, which is a direct
generalization of the eq.(16) from $~n_e=4~$ to
any arbitrary $~n_e\geq 4$.
Only the one-loop graphs from the $~E^2$-level operator
$~(f_\pi^2/4)\,Tr[(D_\mu U)^\dagger(D^\mu U)]~$ and the
tree graphs from the $~E^4$-level
operators \cite{ewcl}, such as
$~\alpha_1(f_\pi/\Lambda)^2 \,[Tr(D_\mu U)^\dagger(D^\mu U)]^2~$ and
$~\alpha_2(f_\pi/\Lambda)^2 \,[Tr(D_\mu U)^\dagger(D^\nu U)]^2~$,
 can contribute to this leading
energy behaviour. The Feynman diagrams from the
other $~E^4$-level operators, such as\footnote{The
custodial $SU(2)$-symmetry violating
operator $~(1/8)\Delta \rho f^2_\pi \,[Tr(\tau^3 U^\dagger
D_\mu U)]^2~$ can contribute to some
pure GB-graphs without internal gauge boson lines,
whose contributions however are at most of
$~~{\cal O}(\Delta\rho\frac{E^2}{f_\pi^2})f_\pi^{D_T}\approx
{\cal O}(\frac{m_t^2}{16\pi^2 f_\pi^2}\frac{E^2}{f_\pi^2})
f_\pi^{D_T}\approx {\cal O}(y_t^2
\frac{E^2}{\Lambda^2})f_\pi^{D_T}\approx {\cal O}(g^2
\frac{E^2}{\Lambda^2})f_\pi^{D_T}~$, where
$y_t$ is the top quark Yukawa coupling.}
\hfill \linebreak
$~-ig\alpha_{9L}(f_\pi/\Lambda)^2 \,
Tr[W^{\mu\nu}(D_\mu U)(D_\nu U)^\dagger]~$,
$~-ig^\prime\alpha_{9R}(f_\pi/\Lambda)^2
 \,Tr[B^{\mu\nu}(D_\mu U)(D_\nu U)^\dagger]~$,~ and
\hfill \linebreak
$~gg^\prime\alpha_{10}(f_\pi/\Lambda)^2 \,
Tr[UB^{\mu\nu}U^\dagger W_{\mu\nu}]~$,
must contain gauge boson lines and are therefore
{\it not sensitive} to probing the SB
sector via longitudinal scatterings. Thus up to $E^4$-level
the condition (21b) gives
$$
{\cal O}(g^2) \ll  \frac{E^2}{f_\pi^2}\frac{E^2} {\Lambda^2}
{}~~,~~~~{\rm or}~~~~~
 \frac{M_W^2}{E^2} \ll \frac{1}{4}\frac{E^2} {\Lambda^2}~.
\eqno(23)                                         
$$

\noindent
We note that the result of (23) holds independent of the
number of external lines involved in pure $~V_L$-scattering processes.
Our condition (18) for
a  pure $~2\rightarrow 2~$ $~V_L$-scattering is only a special case
of (23).

As $E \geq {\cal O}(1)\,$TeV, eq.(23) is satisfied.
Our above precise formulation of the ET, eqs.(21) and (21a,b,c), therefore
provides a rigorous theoretical reasoning for justifying
many previous applications
of the ET in the literature to study the
strongly coupled SB sector by ignoring all the internal
gauge boson lines in the GB-amplitudes.
Most importantly, our result (23) shows that {\it in order
to  probe strongly coupled~~{\rm SB} sector from pure
longitudinal weak-boson scattering processes with
any number of external lines, we must experimentally
measure their production rates in the energy region above $1\,${\rm TeV}. }

\vspace{0.4cm}
\noindent
{\bf 5. Conclusions}

We have examined the Lorentz non-invariance ambiguity
for longitudinal weak-boson scatterings and derived the precise
conditions, eqs.(10a) and (10b) (~or (15)~), for the equivalence of
the $~V_L$-amplitude and the GB-amplitude, as shown in (10).
After analyzing the intrinsic connection between the ET and the problem of
probing the electroweak SB sector, we presented  the universal
formulation of the ET in eqs.(10) and (10a,b) for both
the SM and the general CLEWT.
We have also defined the {\it safe} Lorentz frames in which the
condition (10a) holds. We gave an explicit example,
$Z_L H \rightarrow Z_L H$, to show that the center-of-mass frame
of this scattering process for a heavy Higgs boson
(~$M_W \ll m_H < E_{\rm CM}$~)
is not a safe frame because the (10a) in this case is not satisfied.
Therefore, in the CM frame the $Z_L H \rightarrow Z_L H$
amplitude cannot be estimated by using the
corresponding GB-amplitude $\pi^0 H \rightarrow \pi^0 H$,
as shown in (13).
We note that the above formulation of the ET not only serves as a
technique tool in simplifying the $~V_L$-amplitude calculation using
the GB-amplitude when the conditions (10a,b) are satisfied,
but, most importantly,
{\it this formulation also discriminates processes which are
not sensitive to probing the electroweak SB sector when (10a) or (10b) fails.}
Furthermore, the condition in eq.(15) determines
 whether the $~V_L$-scattering process of interest is sensitive to probing
the SB sector to the desired precision in perturbative calculations.
The minimum energy scale required for testing the SB sector (~assuming no
light resonance present~) of
the SM and the CLEWT beyond the leading order
(~up to $E^4$-level~)
were given in (18) or (23).
We found that longitudinal weak-boson scatterings can only be sensitive
to probing strongly coupled electroweak SB sector in the TeV region, i.e.
$~E\geq {\cal O}(1)\,$TeV.  In this case, for
pure longitudinal weak-boson scatterings, the ET  takes a very simple
form in which the GB-amplitude is calculated by ignoring all the internal
gauge-boson lines and fermion loops (~cf. (21,21a,b,c)~).
Here the multiplicative modification
factors can be exactly simplified as unity in a very simple renormalization
scheme, {\it Scheme-III} ~(~cf. (22)~).

\vspace{1.2cm}
\noindent
{\bf Acknowledgements}\\
H.J.H. is supported in part by
the U.S. Department of Energy under grant DEFG0592ER40709;
Y.P.K. is supported by the National Natural Science Foundation of China and the
Fundamental Research Foundation of Tsinghua University, and would like to
thank the DESY Theory Group for hospitality;
C.P.Y. is supported in part by NSF under grant PHY-9309902.
H.J.H. and C.P.Y. are
grateful to W.W. Repko and Y.P. Yao for helpful discussions.
H.J.H. thanks Lay Nam Chang and Chia Tze for many invaluable discussions,
and the LBL Theory Group for hospitality;
he also thanks Mike Chanowitz for reading the manuscript and
for his kind discussions and suggestions.


\newpage
\vspace{0.3cm}
\thebibliography{References}
\bibitem{cl}
 J.M. Cornwall, D.N. Levin, and G. Tiktopoulos, Phys. Rev. {\bf D10}
(1974) 1145;\\
C.E. Vayonakis, Lett. Nuovo {\bf 17} (1976) 383;\\
B.W. Lee, C. Quigg, and H. Thacker, Phys. Rev. {\bf D16} (1977) 1519.
\bibitem{cg}
M.S. Chanowitz and M.K. Gaillard, Nucl. Phys. {\bf B261} (1985) 379;\\
G.J. Gounaris, R. K$\ddot{o}$gerler, and H. Neufeld, Phys. Rev. {\bf D34}
(1986) 3257;\\
H. Veltman, Phys. Rev. {\bf D41} (1990) 2294.
\bibitem{yy}
Y.-P. Yao and C.-P. Yuan, Phys. Rev. {\bf D38} (1988) 2237;\\
J. Bagger and C. Schmidt, Phys. Rev.  {\bf D41} (1990) 264.
\bibitem{hkla}
H.-J. He, Y.-P. Kuang, and X. Li, Phys. Rev. Lett. {\bf 69} (1992) 2619.
\bibitem{hklb}
H.-J. He, Y.-P. Kuang, and X. Li, Phys.  Rev. {\bf D49} (1994) 4842.
\bibitem{hklc}
H.-J. He, Y.-P. Kuang, and X. Li, Phys. Lett. {\bf B329} (1994) 278.
\bibitem{do}
A. Dobado and J.R. Pelaez, Phys. Lett. {\bf B329} (1994) 469
  ( Addendum, ibid, {\bf B335} (1994) 554 );
                           Nucl. Phys. {\bf B425} (1994) 110.
\bibitem{kilgore}
W.B. Kilgore, Phys. Lett. {\bf B294} (1992) 257.
\bibitem{nambu}
Y. Nambu, {\it  New Theories in Physics }, Proc. XI Worsow Symposium on
Elementary Particle Physics, ed. by Z. Ajudk et al.,
World Scientific, Singapore, 1989, p.1.
\bibitem{jw}
M. Jacob and G.C. Wick, Ann. Phys. (N.Y.) {\bf 7} (1959) 404.
\bibitem{ger}
C. Grosse-Knetter, I. Kuss, BI-TP 94/10; C. Grosse-Knetter,
BI-TP 94/25; (unpublished).
\bibitem{ey}
M. Einhorn, Nucl. Phys. {\bf B246} (1984) 75;\\
R.S. Chivukula and M. Golden, Phys. Lett. {\bf B267} (1991) 233;\\
R.S. Chivukula, M. Golden, and M.V. Ramana,
 Phys. Lett. {\bf B293} (1992) 400;\\
S. Nachulich and C.-P. Yuan, Phys. Lett. {\bf B293} (1992) 395;
                          Phys. Rev. {\bf D48} (1993) 1097.
\bibitem{power}
H.-J. He, Y.-P. Kuang, and C.-P. Yuan,
{\it Generalized Power Counting for Electroweak Theories and Applications to
Probing Symmetry Breaking Sector }, VPI-IHEP-94-09, in preparation.
\bibitem{georgi}
A. Manohar and H. Georgi, Nucl. Phys. {\bf B234} (1984) 189;\\
H. Georgi, Phys. Lett. {\bf B298} (1993) 187.
\bibitem{low}
M. Chanowitz, M. Golden, and H. Georgi, Phys. Rev. Lett. {\bf 57} (1986) 2344;
Phys. Rev. {\bf D36} (1987) 1490.
\bibitem{chano}
M.S. Chanowitz, private communication. See also eq.(1) in\\
M.S. Berger and M.S. Chanowitz, Phys. Rev. Lett. {\bf 68} (1992) 757.
\bibitem{dw}
For examples, \\
S. Dawson and S. Willenbrock, Phys. Rev. {\bf D40} (1989) 2880;\\
M.J.G. Veltman, F.J. Ynduarin, Nucl. Phys. {\bf B325} (1989) 1.
\bibitem{hunter}
For example,  J.F. Gunion, H.E. Haber,
G.L. Kane, and S. Dawson, {\it The Higgs Hunter's Guide },
Addison-Wesley, 1990.
\bibitem{bz}
For a recent review, see,  \\
S.J. Brodsky and P.M. Zerwas,
{\it High Energy Photon-Photon Collisions}, SLAC-PUB-6571, hep-ph/9407326,
1994.
\bibitem{ewcl}
For these non-linear operators, see, for example,\\
 A.F. Falk, M. Luke and E.H. Simmons,
Nucl. Phys. {\bf B365} (1991) 523;\\
J. Bagger, S. Dawson and G. Valencia,  Nucl. Phys. {\bf B399} (1993) 364;
and the references therein.
\end{document}